\documentclass[12pt]{article}
\usepackage[dvips]{graphicx}
\usepackage{pdproc}

  \makeatletter 
  \def\@cite#1{[#1]} 
  \makeatother    
\begin{document}

\renewcommand{\thefootnote}{\alph{footnote}}

\title{ Determination of the $HWW$ and $HZZ$ Couplings at the LHC
\footnote{Argonne report ANL-HEP-CP-04-79. To be published in the 
Proceedings of the 12th International Conference on Supersymmetry 
and Unification of Fundamental Interactions (SUSY 2004) 
June 17-23, 2004, Tsukuba, Japan}
}

\author{ EDMOND L. BERGER}

\address{ 
High Energy Physics Division, Argonne National Laboratory \\
Argonne, Illinois 60439, USA
\\ {\rm E-mail: berger@anl.gov}}

\abstract{
The weak boson fusion process for neutral Higgs boson production is 
investigated with particular attention to the accuracy with which the 
Higgs boson couplings to weak bosons can be determined in final states 
that contain a Higgs boson plus at least two jets at CERN Large Hadron 
Collider energies. We determine that an accuracy of 
$\delta g/g \sim 10$\% on the effective coupling $g$ may be possible after 
the accumulation of $\sim 200$~fb$^{-1}$ of integrated luminosity.  
}

\normalsize\baselineskip=15pt

\section{Introduction and Motivation} 

Following the discovery of the neutral Higgs boson $H$ at the Fermilab 
Tevatron or the CERN Large 
Hadron Collider~(LHC), attention will focus on the measurement of its 
couplings to gauge bosons and fermions.  A promising reaction from which
to extract some of these couplings, particularly the $HWW$ coupling, is 
the weak-boson fusion (WBF) process, in which the Higgs boson is produced 
via fusion of the weak bosons $W$ and $Z$: $WW, ZZ \to H$, and is 
accompanied in the final state by two jets that carry large transverse 
momentum $p_T$.  In this paper, I summarize a recent study done in 
collaboration with John Campbell in which we simulate $H +2$~jet events 
and investigate how well the Higgs boson couplings $g$ to the weak 
vector bosons, $W$ and $Z$, can be determined~\cite{Berger:2004pc}. 

We assume that a standard-model-like Higgs boson has been discovered 
with mass in the range $115 < m_H < 200$~GeV, and that a sample exists of 
$H +2$~jet events at the LHC. We focus on two production subprocesses that 
contribute to the $H +2$~jet event sample: (1) the WBF signal subprocess 
$W + W \rightarrow H + X$ and $Z+ Z \rightarrow H +X$, and (2) the 
irreducible QCD background subprocess, {\it e.g.}, $g + g \rightarrow H +X$.  
Once a sample exists of $H +2$~jet events, the salient issue for the 
determination of couplings is this: How well can we resolve WBF production 
of $H$ from QCD production of $H$?  

In our work, we perform an independent calculation of the signal and 
background $H +2$~jet processes to gauge the effectiveness of 
cuts used to select the WBF signal and to evaluate the accuracy with which 
the coupling $g$ can be determined.  
We define signal ``purity'' as $P = S/(S+B)$ where $S$ is the number 
of signal $H+2$~jet events and $B$ is the number of $H+2$~jet QCD 
background events, both in the WBF region of phase space. We study $P$ 
of the signal {\it vs.} the cut on the jet transverse momentum 
$p^{\rm cut}_T$ used to define the event sample.  We evaluate the 
uncertainty $\delta g /g$ of the coupling in terms of $P$, the experimental 
statistical uncertainty of the event sample $\delta N /N$, and the estimated theoretical 
uncertainties of the signal $\delta S /S$ and of the background  
$\delta B /B$.

\section{Signal and Background; Calculations and Uncertainty Estimates}

The WBF $H+2$~jet signal region is characterized by jets that carry large 
transverse momentum. Correspondingly, hard QCD matrix elements must be 
used in order to represent the signal and the $H+2$~jet 
background reliably. We use the dipole subtraction method to compute the 
NLO corrections to the WBF $H+2$~jet cross section in a fully differential 
fashion in order to examine in detail the effects of the WBF selection cuts. 
Our independent calculation verifies the NLO results of Ref.~\cite{Figy:2003nv}. 
The NLO $K$-factor is modest $\sim 10$\%, and it varies but slightly over the 
phase space appropriate for the WBF signal. The hard perturbative scale $\mu$ 
dependence of the signal cross section is also modest, showing a $\pm 2$\% change 
over the range $\frac{1}{2}m_H < \mu < 2 m_H$. An additional $\pm 3$\% 
uncertainty may be attributed to parton density variation, based on the CTEQ6 
procedure for estimating such uncertainties. The WBF $H+2$~jet signal process 
can therefore be assigned a relatively small theoretical uncertainty. We use 
$\delta S /S \simeq 5$\%.  

We also use fully differential perturbative QCD expressions for the 
background $H+2$~jet matrix elements.  At present, these background 
distributions are known only at leading order~\cite{Kauffman:1996ix}. The 
LO $\mu$ dependence is substantial~\cite{Berger:2004pc} in the WBF region 
of phase space.  The cross section is $\sim 70$\% greater at 
$\mu = \frac{1}{2}m_H$ and $\sim 40$\% less at $\mu = 2m_H$ than at 
$\mu = m_H$. The NLO $K$-factor may be estimated from NLO calculations 
of the {\it inclusive} rate~\cite{Harlander:2001is}, $K \sim 1.7 - 1.8$, 
and the NLO $H+1$~jet cross section~\cite{Ravindran:2003um}, 
$K \sim 1.3 - 1.5$. At NLO, the $\mu$ dependence is less, $\sim 20$\%, 
and the PDF uncertainty is another $5$\%. For the uncertainty $\delta B /B$, 
we adopt the NLO estimates of $\mu$ dependence and PDF uncertainty, despite 
the fact that the NLO $H+2$~jet calculation is not complete.  We use 
$\delta B /B \simeq 30$\% in our subsequent estimation of $\delta g /g$.

\section{Event Characteristics}

The hallmark of WBF events in hadron reactions is a Higgs boson 
accompanied by two ``tagging'' jets having large 
$p_T \sim {\cal O}(\frac{1}{2} M_W)$. The QCD 
$gg \rightarrow H + 2$~jets background process generates a softer 
$p_T$ spectrum.  The rapidity spectra for the WBF and QCD production 
mechanisms also differ, related to the fact that the gluon parton density 
(that plays a dominant role in generating the background) is softer than 
the quark density that drives the WBF signal. Motivated by our comparison 
of rapidity spectra~\cite{Berger:2004pc}, we choose a uniform rapidity cut 
to define our WBF $H + 2$~jet sample, a cut that ensures at least one jet 
lies within the peak of the rapidity distribution of the WBF signal: 
\begin{equation}
\eta_{\rm peak}-\eta_{\rm width}/2 < |\eta_{j}| <
 \eta_{\rm peak}+\eta_{\rm width}/2,
\label{etapeak}
\end{equation}
for $j=j_1$ or $j=j_2$, where $\eta_{\rm peak}$=3 and
$\eta_{\rm width}$=2.8. This simple definition of the WBF sample offers 
advantages in a high luminosity environment where a large value of 
$p_T^{\rm cut}$ is appropriate and multiple events per crossing may 
be an issue.

Having defined our signal region, we compute event rates for the 
$H + 2$~jet WBF signal and QCD background processes, and we compute 
the signal purity $P = S/(S+B)$. A $p_T$ cut of $20$~GeV is barely 
sufficient to distinguish the WBF signal above the QCD LO background 
for $m_H=115$~GeV. The signal $S$ to background $B$ ratio improves to 
about 2 for $p^{\rm cut}_T \ge 40$~GeV. At $m_H=200$~GeV, a $S/B$ of 
about $1.7$ is obtained for $p^{\rm cut}_T \sim 20$~GeV, rising to 
$\sim 3$ for $p^{\rm cut}_T \ge 40$~GeV. We determine that a $p_T$ cut 
of $40$~GeV yields a good $S/B$ across the Higgs boson mass range 
$m_H=115$--$200$~GeV. 
Signal purities of $\sim 65$\% are obtained for $p^{\rm cut}_T \ge 40$~GeV; 
purity is greater at the larger values of $m_H$.  

\section{Coupling Uncertainties}

Both the signal $S$ and the background $B$ have $H + 2$~jets. The total 
number of events is $N = S + B$. We derive an equation for 
the uncertainty in the effective $HWW$ coupling $g$ in terms of purity $P$:
\begin{equation}
\delta g /g = 1/2 \sqrt {[ (\delta S/S)^2 + (1/P)^2(\delta N/N)^2 + ((1-P)/P)^2(\delta B/B)^2 ]} .
\label{uncert}
\end{equation}
\noindent A minimum of $\sim 10$~fb$^{-1}$ in integrated luminosity is 
needed to discover the Higgs boson in the WBF process~\cite{Asai:2004ws}, 
corresponding to  one year of LHC operation at 
$10^{33}$ cm$^{-2}$s$^{-1}$.  The values of $N$ and $\delta N/N$ depend on 
the specific decay modes of the Higgs boson.  For $m_{H} = 115$~GeV, we 
pick $H \rightarrow \tau^+ \tau^-$, with one $\tau$ decaying to hadrons and 
one to leptons. For $m_H = 200$~GeV, we use $H \rightarrow W^+ W^-$, with 
both $W$'s decaying to leptons.  With $p_T^{\rm cut} = 40$~GeV, we find 
$\delta N/N \sim 10$\% at $m_H = 115$~GeV in the $H \rightarrow \tau^+ \tau^-$ 
mode and $\delta N/N \sim 6$\% at $m_H = 200$~GeV in the $W^+ W^-$ decay mode.  

After 5 years of LHC operation, we may anticipate an integrated luminosity 
of $200$~fb$^{-1}$. The corresponding values are 
$\delta N/N \sim2$\% in the $\tau \tau$ mode at $m_H = 115$~GeV,  
and $\delta N/N \sim 1.5$\% at $m_H = 200$~GeV in the $WW$ mode, both 
for $p_T^{\rm cut} = 40$~GeV.  

\begin{figure}[h]
\begin{center}
\includegraphics*[angle=270,width=10cm]{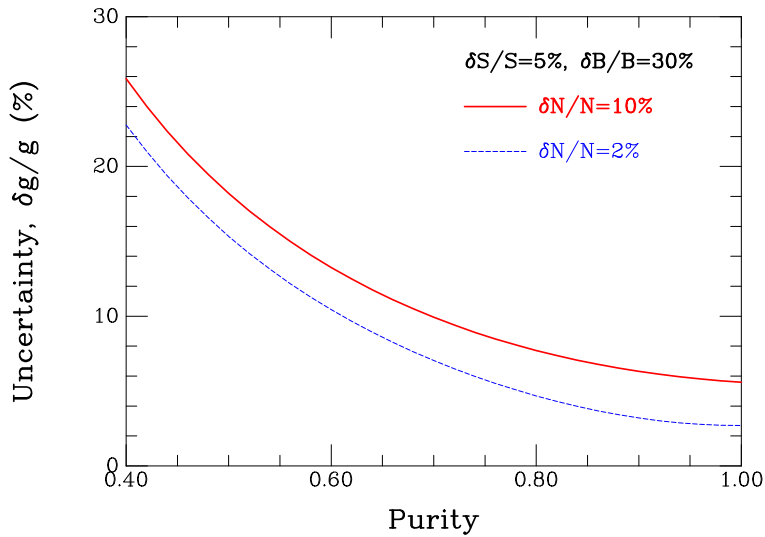}
\caption{The predicted uncertainty $\delta g/g$ in the effective coupling of 
the Higgs boson to a pair of $W$ bosons is shown as a function of signal 
purity $P = S/(S+B)$ for expected statistical accuracies $\delta N/N$ of 
10\% and 2\%, and uncertainties in knowledge of the signal $S$ and 
background $B$ of 5\% and 30\% respectively.}
\label{uncert}
\end{center}
\end{figure}

In Fig.~\ref{uncert}, we show our calculation of the expected uncertainty 
$\delta g/g$ as a function of purity $P$ for both high and low luminosity 
samples at the LHC. If $\delta N/N \sim 10$\%, we find $\delta g/g \sim 10$\% 
for $P =0.7$.  With $\delta N/N \sim 2$\%, $\delta g/g \sim 7$\% for $P =0.7$.  
The uncertainties in $S$ and in $B$ dominate uncertainty in $g$.  With $P = 0.7$ 
and $\delta N/N = 2$\%, $\delta S /S$ and $\delta B/B$ would have to be reduced 
to $3$\% and $6$\% before statistics control the answer. We conclude that $P > 0.65$ 
permits $\delta g/g \sim 10$\% after $200$~fb$^{-1}$, obtained for 
$p_T^{\rm cut} > 40$~GeV at $m_H = 115$~GeV and for $p_T^{\rm cut} > 20$~GeV at 
$m_H = 200$~GeV.  These estimates assume a LO value for $B$.  If we suppose 
$K^{\rm NLO}_{\rm background} \sim 1.6$, then we find $P = 0.56$ at $m_H = 115$~GeV 
for $p_T^{\rm cut} > 40$~GeV, and $\delta g/g$ rises to $13$\%. At $m_H = 200$~GeV,  
the new values would be $P = 0.52$ for $p_T^{\rm cut} > 20$~GeV, and 
$\delta g/g = 15$\%.  

It may be suggested that greater purities and accuracies could be 
achieved if one of the alternative definitions of the WBF sample is used. 
Using a cut on the rapidity separation between the two tagging jets favored, 
{\it e.g.}, in Ref.~\cite{Figy:2003nv}, we find that the signal rate is 
diminished somewhat and that the purity is greater.  However, the quantitative 
shift from $P = 0.67$ to $P = 0.78$ at $M = 115$~GeV and $p_T > 40$~GeV is an 
improvement of only $3$\% in $\delta g/g$, and this reduction is offset somewhat 
by the loss in statistical accuracy.  

Our WBF signal purity and our uncertainties 
are obtained in a well controlled situation in which there is an identified 
Higgs boson in a sample of $H + 2$~jet events. In an experiment, there 
will be additional sources of background from final states that mimic a Higgs 
boson, the effects of which presumably only increase the expected uncertainties.  

We conclude that after $200$~fb$^{-1}$ are accumulated at the LHC, it may be 
possible to achieve an accuracy $\delta g/g \sim 10$\% in the effective coupling 
(combination of $HWW$ and $HZZ$ couplings) of the Higgs boson to weak bosons. 
These estimates are about a factor of 2 less optimistic than those in the 
Les Houches 2003 study~\cite{Duhrssen:2004cv}. The salient difference may be 
traced to assumptions in Ref.~\cite{Duhrssen:2004cv} about the size of the 
irreducible $H + 2$~jet QCD background. In order to reduce the estimated 
uncertainty in $g$, the next major step would be a fully differential NLO 
calculation of the $H + 2$~jet backgrounds.

\section{Acknowledgments}

This work was supported by the U.~S.~Department of Energy under contract
No.~W-31-109-ENG-38. 

\bibliographystyle{plain}

\end{document}